\documentclass[reprint,aps,prb,amsmath,amssymb]{revtex4-1}

\usepackage[]{graphicx}
\usepackage[]{units}
\usepackage{times}
\usepackage{bm}
\usepackage[ulem=normalem]{changes}
\usepackage[normalem]{ulem}		
\usepackage{color}
\usepackage{xr} 				
\usepackage{cancel}
\usepackage{hyperref}
\usepackage{babel}
\definechangesauthor[color=blue]{SM}

\renewcommand{\emph}{\textit}

\newcommand{\one}{1\!\!1}

\begin{document}

\title{\textit{Ab initio} Maxwell-Bloch Approach 
for X-Ray Excitations in Two-Dimensional Materials }
\author{Joris Sturm$^{1}$, Ivan Maliyov$^{3}$, Dominik Christiansen$^{1,2}$, Malte Selig$^1$, Marco Bernardi$^2$, Andreas Knorr$^1$}
\affiliation{$^1$ Technische Universität Berlin, Institut für Physik und Astronomie, Nichtlineare Optik und Quantenelektronik, 10623 Berlin, Germany}
\affiliation{$^2$ Department of Applied Physics and Materials Science, and Department of Physics, California Institute of Technology, Pasadena, California 91125, USA}
\affiliation{$^3$ Mathematics for Materials Modelling, Institute of Mathematics and Institute of Materials, EPFL, CH-1015 Lausanne, Switzerland}

\begin{abstract}
The combination of Maxwell and X-ray Bloch equations forms an appropriate framework to describe ultrafast time-resolved X-ray experiments on attosecond time scale in crystalline solids. However, broadband experiments such as X-ray absorption near edge spectroscopy or resonant inelastic X-ray scattering require a detailed knowledge of the electronic structure and transition matrix elements. Here, we show how to fill this gap by combining the Maxwell-X-ray Bloch formalism with first-principles calculations treating explicitly the core states. The resulting X-ray absorption spectrum recovers key spectral signatures which were missing in our previous work \cite{christiansen2022theory} relying on a semi-empirical tight-binding approach.
\end{abstract}

\maketitle


\section{Introduction}

Recent advances in laser technology have enabled the creation of attosecond short pulses in the extreme ultraviolet to soft X-ray regime \cite{drescher2001x,cousin2017attosecond,duris2020tunable,xu2020production}. 
In turn, this allows one to study and even control coherent effects on ultrashort timescales below the femtosecond limit. Such attosecond dynamics was initially studied in atomic systems, particularly for real-time observation of valence electron motion \cite{goulielmakis2010real}. These investigations were quickly followed by studies on molecular systems \cite{calegari2014ultrafast,nisoli2017attosecond,peng2019attosecond,palacios2020quantum}, where light-induced ultrafast dynamics plays a role in early photosynthetic processes \cite{young2016structure,kern2018structures} and in chemical reactions or interface dynamics involving electron transfer \cite{kasai2018x,bag2021attochemistry,gretendipolar}. 
In the few last years, attosecond dynamics in crystalline solids has also become widely studied. Compared to femtosecond dynamics, attosecond short X-ray excitation in solids is still an unexplored area of research. A key experimental technique is attosecond transient absorption, which combines time-delayed pump and probe pulses. These advances in attosecond experiments call for theoretical approaches accompanying the experimental efforts. 
Density functional theory (DFT) has enabled major progress in modeling X-ray processes~\cite{rehr2000theoretical,bunuau2012time}. Yet, non-equilibrium effects are beyond the reach of DFT, and extensions such as time-dependent DFT struggle with the description of coherent effects occurring at overlapping pump and probe pulses or at subfemtosecond time scale due to its high computational cost. 
\begin{figure}[t]
\begin{center}
\includegraphics[width=\linewidth]{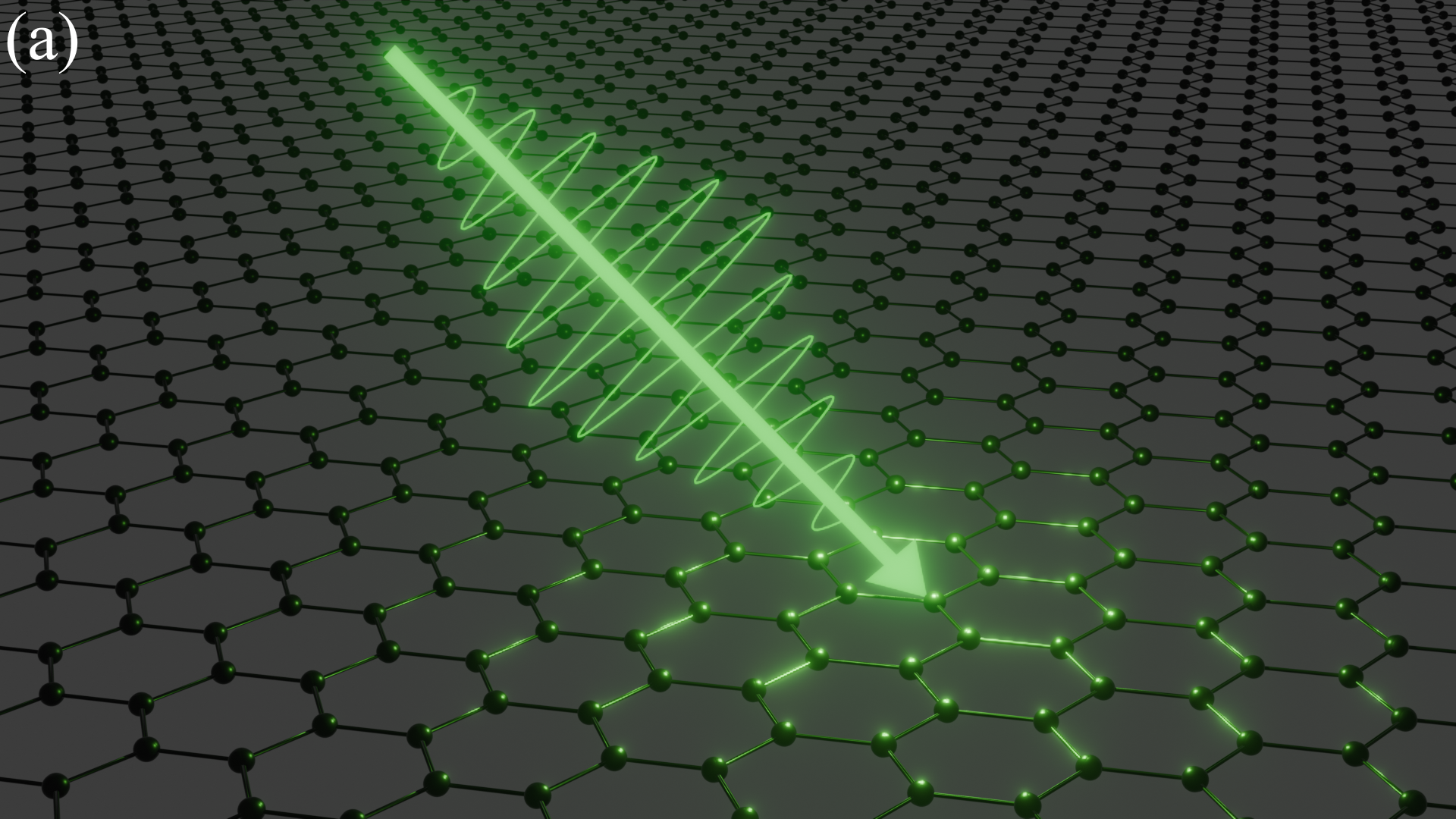}
\includegraphics[width=\linewidth]{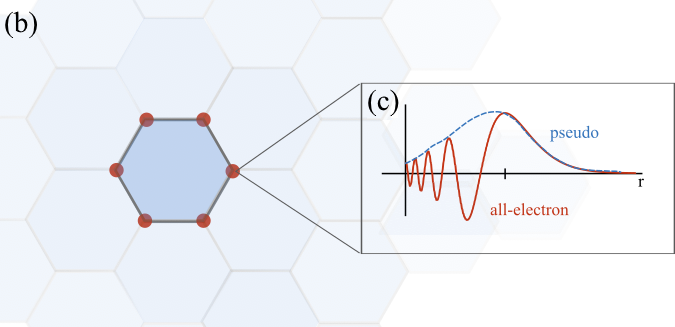}
\end{center}
\caption{(a) Interaction of the X-ray with the 2d crystal lattice (b) Sketch of a hexagonal crystal lattice in real space. (c) The zoom-in on an atom shows the tightly bound core-electrons. The wave functions of the core electrons oscillate rapidly close to the nucleus due to the divergent Coulomb potential, while the pseudopotential model gives a smoother wave function that takes only the envelope into account. }
\label{fig:sketch}
\end{figure}

To address nonequilibrium X-ray dynamics, in our recent work we proposed a self-consistent Maxwell-X-ray Bloch equation approach and applied it to X-ray excitations in two-dimensional materials \cite{christiansen2022theory}. This approach enables the description of key features in nonlinear X-ray experiments, including aspects of many-body interaction such as energy renormalization, valence or core excitons, and relaxation dynamics. This method allows us to describe the real-time dynamics starting from the pump pulse and tracking the dynamics up to the picosecond time scale. 
Due to the self-consistent coupling between the dynamical X-ray Bloch equations and the Maxwell equations, our formalism can predict experimentally observable signals as well as capture X-ray propagation in the material, localized excitation of the sample, and polaritonic effects. As proof of principle, in our recent work we calculated X-ray absorption spectra in graphene including X-ray absorption near edge spectroscopy (XANES) and extended X-ray absorption fine structure (EXAFS), all within a single calculation \cite{christiansen2022theory}. 
However, our previously developed approach also suffers from several limitations. First, as evidenced by comparing with experiments, our predicted spectrum missed a key spectral feature. The reason for this discrepancy is that our analytical tight-binding approach included only a small number of bands, while a broadband experiment such as XANES requires a detailed knowledge of the full band structure. Second, for reliable predictions of the dynamics it is important to compute the interaction matrix elements with quantitative accuracy. This requires full access to the Bloch electronic wave functions (or equivalently, to the tight-binding coefficients).

In this work, we show an approach combining first-principles calculations with the Maxwell-X-ray Bloch formalism to address both of these challenges. We compute the core, valence and conduction band states on equal footing using DFT calculations with the \textsc{Quantum ESPRESSO} package~\cite{giannozzi2009quantum}, and then use these quantities as a starting point to obtain accurate interaction matrix elements, focusing on transition dipoles computed with the YAMBO code~\cite{sangalli2019many}. This  allows us to develop a combined Maxwell-X-ray Bloch plus first-principles scheme to reconstruct the full XANES spectrum. Our combined approach paves the way for a more accurate and general description of X-ray-matter interactions and X-ray excitation dynamics in two-dimensional materials.

\section{Band energies and transition dipoles from first principles}

We develop a theoretical description of X-ray interactions with two-dimensional (2D) crystals. In crystalline solids, because of the periodic lattice potential the electronic wave functions satisfy Bloch's theorem. 
The electronic wave function for band $n$ and crystal momentum $\mathbf{k}$ reads
\begin{align}
    \psi_{n,\mathbf{k}}(\mathbf{r})=\frac{1}{\sqrt{Al_z}}e^{i\mathbf{k\cdot r}}u_{n,\mathbf{k}}(\mathbf{r}) \label{eq:Bloch}, 
\end{align}
where $A$ is the area of the 2D material and $l_z$ is the size of the simulation cell in the layer-normal direction. The Bloch wave functions are products of a plane wave and a lattice periodic function $u_{n\mathbf{k}}(\mathbf{r})$, with in-plane wave-vector $\mathbf{k}$ and position $\mathbf{r}$.
\\
\indent
Using plane-wave basis, the Bloch functions are expressed as a Fourier expansion with respect to the reciprocal lattice vectors $\mathbf{G}$,
\begin{align}
    u_{n\mathbf{k}}(\mathbf{r})=\sum_{\mathbf{G}}c_{n\mathbf{k+G}}e^{i\mathbf{G\cdot r}} \label{eq:BlochU} \;.
\end{align}
Inserting Eqs. \eqref{eq:Bloch} and \eqref{eq:BlochU} into the Schr\"odinger equation allows one to solve for the expansion coefficients $c_{n\mathbf{k,G}}$. In DFT, these calculations are based on the Kohn-Sham framework~\cite{hohenberg1964inhomogeneous,kohn1965self}, which describes a system of independent electrons in an effective potential written as the sum of nuclei, Hartree and exchange-correlation contributions~\cite{Martin2020}. 
\\
\indent
Typical DFT calculations treat explicitly only the valence electrons and leave out the core electrons due to their negligible contribution to chemical bonding and optical properties of the material. The effect of the core electrons is included by using pseudopotentials~\cite{Martin2020}, which approximate interactions between the valence electrons and the nuclei plus core electrons~\cite{pickett1989}. 
%
Figure~\ref{fig:sketch}(a) shows an exemplary hexagonal lattice with two atoms per unit cell. 
Figure \ref{fig:sketch}(b) compares the wave functions of a valence electron obtained from an all-electron calculation versus one using pseudopotentials. The oscillatory behavior of the all-electron wave function near the nucleus stems from the constraint to be orthogonal to the lower-energy core states with the same angular momentum. Accurately describing these oscillations requires a large kinetic energy cutoff in a plane-wave basis. 
In contrast, the pseudo wave function is smooth in the core region and it accurately reproduces the all-electron wave function outside a cutoff radius from the nucleus in the region relevant for chemical bonding. 
Therefore, pseudopotential calculations cannot be employed when an explicit description of the core electrons is needed, as is the case when modeling X-ray spectroscopies. 
\begin{figure}[t]
\begin{center}
\includegraphics[width=\linewidth]{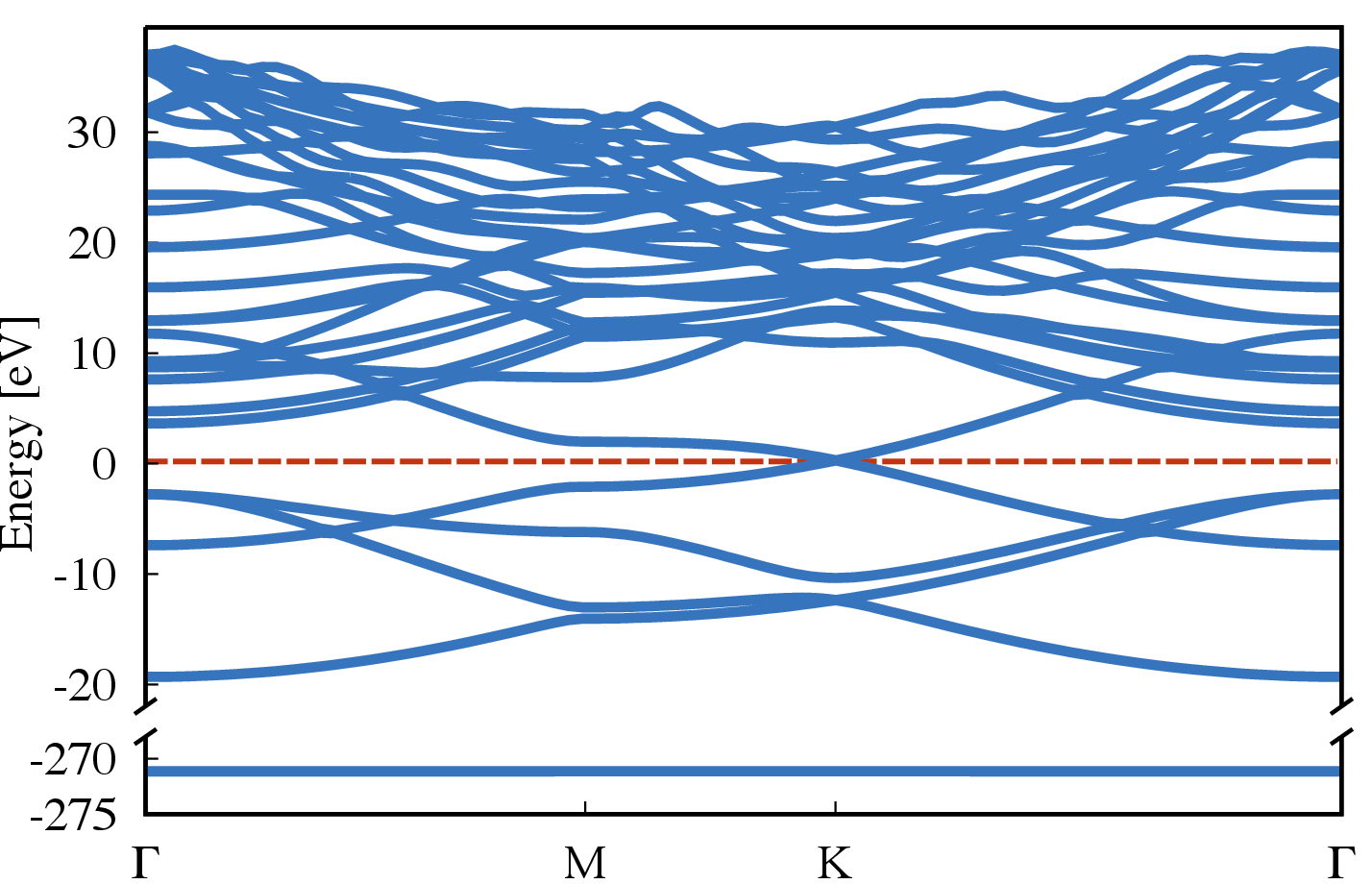}
\end{center}
\caption{Band structure of graphene obtained from a DFT that includes the 1$s$ core state. The dashed line indicates the Fermi level.}
\label{fig:bands}
\end{figure}

For the self-consistent field calculation we describe the atomic species with core-electrons by the projected augmented wave 
(PAW) method \cite{blochl1994projector} as implemented in \textsc{Quantum ESPRESSO} \cite{giannozzi2017advanced}. PAW unifies accuracy of all-electron calculations with the efficiency of pseudopotential methods. This is achieved by a linear transformation which ensures that the pseudo-wavefunction and the all-electron wavefunction differ only in the vicinity of atomic nuclei, in an augmentation region. The transformation is defined using a set of all-electron partial waves, pseudo partial waves, and projector functions. Consequently, the PAW method effectively transforms kinks and rapid oscillations of valence wavefunctions near ion cores into smooth wavefunctions, which are more computationally convenient. Finally, using a generated PAW potential with the desired number of core electrons, and solving the corresponding transformed Kohn-Sham equation, yields Kohn-Sham wave functions in reciprocal space for each $\mathbf{k}$-value and band including core bands, cf. Eq. \eqref{eq:BlochU}. 
Figure \ref{fig:bands} shows a calculated band structure of graphene including the 1$s$ core band along a high-symmetry $\mathbf{k}$-path. The calculated core band energy aligns well with values present in literature \cite{susi2015calculation,xu2015x}. We use graphene as an exemplary material to test the combined \textit{ab initio} X-ray Bloch approach. The PAW potential is generated using the Perdew–Burke–Ernzerhof (PBE) functional with a matching radius (rcore parameter) of \unit[0.7]{a.u.} for the smoothing of the core charge. The electronic \textit{ab initio} calculation is performed with a cutoff energy of \unit[120]{Ry} on a 10x10x1 grid.

\begin{figure}[t]
\begin{center}
\includegraphics[width=\linewidth]{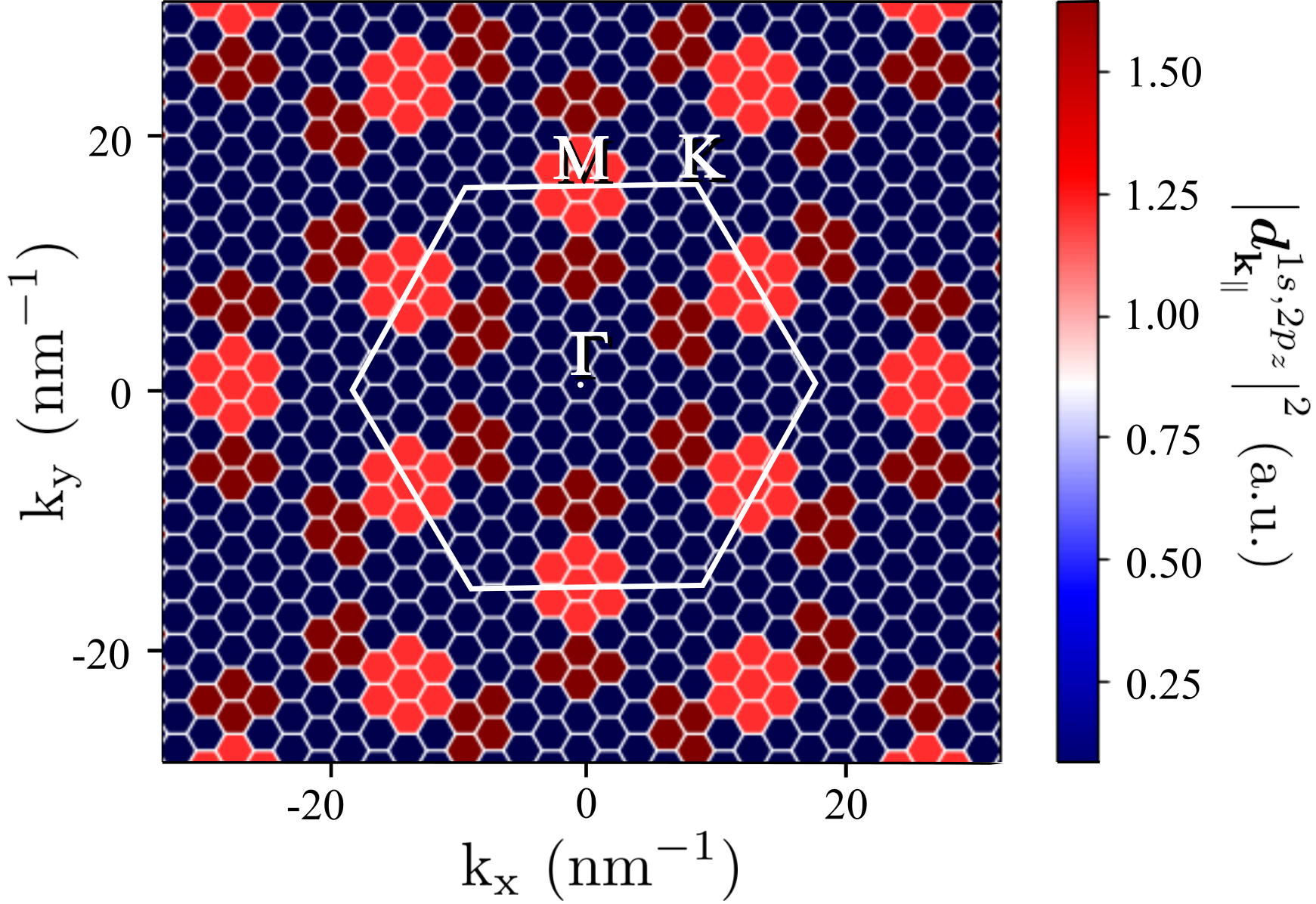}
\end{center}
\caption{Out-of-plane component of the dipole matrix element of graphene for the 1$s$ to $\pi$ (2$p_z$) transition. The matrix element carries the hexagonal symmetry and displays considerable oscillator strength around the M point and on the trace to the $\Gamma$ point. The strength of dipole matrix element is displayed in arbitrary units. }
\label{fig:dipole}
\end{figure}

As exemplary interaction Hamiltonian, we investigate a semi-classical light-matter interaction and evaluate it explicitly with the \textsc{YAMBO} code \cite{sangalli2019many}. The light-matter interaction is described by
\begin{align}
    H=-\frac{1}{Al_z}\sum_{n,\mathbf{k,Q,G}} \mathbf{d}^{nm}_{\mathbf{k+Q,k}}\cdot\mathbf{E}_{-\mathbf{Q+G}}(t) a^{\dagger}_{n,\mathbf{k+Q}}a^{\mathstrut}_{m,\mathbf{k}} \label{eq:light-matter Hamilton}
\end{align}
with an electron creation $a^{\dagger}_{n,\mathbf{k+Q}}$ in band $n$ with wave vector $\mathbf{k+Q}$ and an electron annihilation $a^{\mathstrut}_{m,\mathbf{k}}$ in band $m$ with wave vector $\mathbf{k}$. The electronic transition is determined by the dipole matrix element $\mathbf{d}^{nm}_{\mathbf{k+Q,k}}=ie_0\int_{UC} d^3r ~ u_{m,\mathbf{k}}(\mathbf{r})\nabla_{\mathbf{k+Q}}u^*_{n,\mathbf{k+Q}}(\mathbf{r})$. Here, $\mathbf{Q}$ corresponds to the Fourier momentum from the in-plane position taking into account the spatial resolution of the X-rays: $\mathbf{E}(\mathbf{r},t)=\sum_{\mathbf{Q}}\mathbf{E}_{\mathbf{Q}}(z,t)\exp(i\mathbf{Q\cdot r}_{\parallel})$. The $\mathbf{Q}=0$ limit corresponds to a perpendicular, spatial homogeneous excitation. The index $m$ ensures that only interband transitions are taken into account. Due to the high frequency of X-rays, intraband transition can be neglected. The $\mathbf{Q+G}$ takes into account for possible Umklapp processes originating from the spatial resolution of the X-ray waves.

For soft X-rays interacting with the light carbon atoms of graphene the transition matrix element can be accurately described within a dipole approximation \cite{christiansen2022theory}. However, a careful check for each materials should be performed, especially when entering the medium X-ray regime. Although, performing a dipole approximation in the matrix element, we keep the off-diagonal indices in the electronic operators of Eq. \eqref{eq:light-matter Hamilton} to account for possible spatial inhomogeneity in the dynamics. We compute the dipole moments with \textsc{YAMBO} \cite{onida2002electronic,del1993optical}, which reads the \textit{ab intio} band energies and wave functions from \textsc{Quantum ESPRESSO}. As example, Figure \ref{fig:dipole} shows the out-of-plane component of the dipole integral for a 1$s$ to $\pi$-conduction band (2$p_z$ orbital) transition, which dominates over the in-plane components due to symmetry reasons. As can be seen in Fig. \ref{fig:dipole}, the integral inherits the hexagonal lattice symmetry of graphene. While the dipole integral vanishes at the K points, the transitions are favourable around the M points and between M and $\Gamma$ point. 

\section{X-ray near edge absorption spectrum of graphene}

The response of a two-dimensional material to a weak X-ray excitation can be investigated by the X-ray absorption spectrum. In order to calculate the X-ray absorption of material, we need to find expressions for the reflected and transmitted X-ray light. The full electric field space and time dependence is given by the wave equation
\begin{align}
    \nabla^2\mathbf{E}(\mathbf{r},t)-\frac{\epsilon}{c^2}\frac{\partial^2}{\partial t^2}\mathbf{E}(\mathbf{r},t)&=\mu_0\frac{\partial^2}{\partial t^2}\mathbf{P}(\mathbf{r},t) \nonumber \\
    &-\frac{1}{\epsilon\epsilon_0}\nabla(\nabla\cdot\mathbf{P}(\mathbf{r},t))
\end{align}
with the dielectric constant $\epsilon_0$, the magnetic permittivity $\mu_0$, the speed of light in vacuum $c$, and the dielectric constant of the uniform background $\epsilon$.  Here, we explicitly keep the spatial resolution of the electric field. The macroscopic polarization $\mathbf{P}(\mathbf{r},t)$ provides a connection to the microscopic polarization $ p^{nm}_{\mathbf{k+Q,k}}$. The microscopic polarization describes the transition amplitude from a state $|m,\mathbf{k+Q}\rangle$ to a state $|n,\mathbf{k}\rangle$. The off-diagonal character of the wave vector indices allows for a spatial localized excitation of the sample by the short wavelength X-ray light. In Fourier space, the macroscopic polarization at the sample position $z=0$ reads $\mathbf{P}_{\mathbf{Q}}(t)=\sum_{n,m,\mathbf{k}} \mathbf{d}^{nm}_{\mathbf{k+Q,k}} p^{nm}_{\mathbf{k+Q,k}}(t)$, where the microscopic polarization can be calculated from the X-ray Bloch equations. A self-consistent solution of the wave equation together with the X-ray Bloch equations yields the necessary electric field containing the response of the material \cite{christiansen2022theory}. Then, the electric field can be found as
\begin{widetext}
\begin{align}
    \mathbf{E}_{\mathbf{Q}+\mathbf{G}}(z,\omega)&=\epsilon_0[G_{\mathbf{Q}+\mathbf{G}}](z,\omega)[\chi_{\mathbf{Q}}](\omega) \left[\one - \epsilon_0\sum_{\mathbf{G}'} [G_{\mathbf{Q}+\mathbf{G}'}](z_0,\omega)[\chi_{\mathbf{Q}}](\omega)\right]^{-1}\sum_{\mathbf{G}''}\mathbf{E}^0_{\mathbf{Q}+\mathbf{G}''}(z_0,\omega) + \mathbf{E}^0_{\mathbf{Q}}(z,\omega) \;, \label{eq:electricfield}
\end{align}
\end{widetext}
where $\mathbf{E}^{0}_{\mathbf{Q}}$ is the incident electric field, the Green's dyade $[G_{\mathbf{Q}}](z,\omega)$ describing the propagation of the X-ray field, and the susceptibility $[\chi_{\mathbf{Q}}](\omega)$. The Green's dyade is defined in the appendix. The first term describes the interference of the electric field with the polarized responding material. The reciprocal lattice vector takes into account the Umklapp processes, due to the X-ray wave vector \cite{mahan2013many}. Moreover, Eq. \eqref{eq:electricfield} can describe X-ray scattering at the sample. After the incidence of the X-ray light with wave vector $\mathbf{Q}$ investigating the transmitted X-ray light as function of the changing X-ray wave vector (or angle) yields an X-ray scattering spectrum. However, in the following we are interested in absorption, where only the $\mathbf{G}=0$ limit contributes. Focusing on the K edge, the initial band can be fixed to 1$s$. From the X-ray Bloch equations, the K-edge susceptibility tensor reads in frequency space
\begin{align}
    [\chi_{\mathbf{Q}}](\omega)=\sum_{n,\mathbf{k}}\frac{\mathbf{d}^{1s, n}_{\mathbf{k,k+Q}}\otimes\mathbf{d}^{n, 1s}_{\mathbf{k+Q,k}}}{\hbar\omega-E^{n}_{\mathbf{k}}+E^{1s}_{\mathbf{k+Q}}+i\gamma} \;. \label{eq:susc}
\end{align}
The $n$-sum runs over all conduction bands. The dipole moments $\mathbf{d}^{1s, n}_{\mathbf{k+Q,k}}$ and energies $E^{n}_{\mathbf{k}}$ were explicitly calculated from first principle as described in the previous two sections. The damping constant $\gamma$ is so far phenomenologically added. It can be microscopically calculated by including Coulomb interaction to the 
X-ray Bloch equations as derived in Ref. \cite{christiansen2022theory}, using the \textit{ab initio} Coulomb matrix elements from \textsc{YAMBO}. A detailed knowledge of the Coulomb matrix elements is of crucial interest for the calculation of Meitner-Auger recombination of core-holes, which is a major contribution to the finite lifetime of core excitations \cite{inhester2013core,ohno1992dynamics}. Here, we simply increase the dephasing constant with increasing conduction band index to simulate the larger amount of possible scattering channels. The X-ray wave vector is connected to the frequency $\omega$ and the angle of incidence $\theta$ of the light by the linear light dispersion via $|\mathbf{Q}|^2=\omega^2\sin(\theta)/c^2$. For $\theta=0^\circ$, we recover the special case of a perpendicular excitation geometry. The transmitted and reflected light spectra are defined as the intensity of the X-ray light behind the sample, and the intensity of the X-ray light in front of the sample propagating away from the sample, respectively. 

Finally, it should be noted that the susceptibility tensor (Eq. \eqref{eq:susc}) does not include core-excitonic effects.
Using the corresponding Coulomb matrix element and solving the Bethe-Salpeter equation yields a core-exciton landscape and wave functions \cite{chen2019ab,chen2020exciton}. With this basis set, the electron-core-hole Coulomb interaction in the Maxwell-X-ray Bloch equations can be diagonalized and the equations expressed in a core exciton basis. This enables access to ultrafast dynamics of core-excitons, which are of growing interest for the X-ray community \cite{geneaux2020attosecond,gaynor2021solid,chang2021coupled}.

\begin{figure}[ht]
\begin{center}
\includegraphics[width=\linewidth]{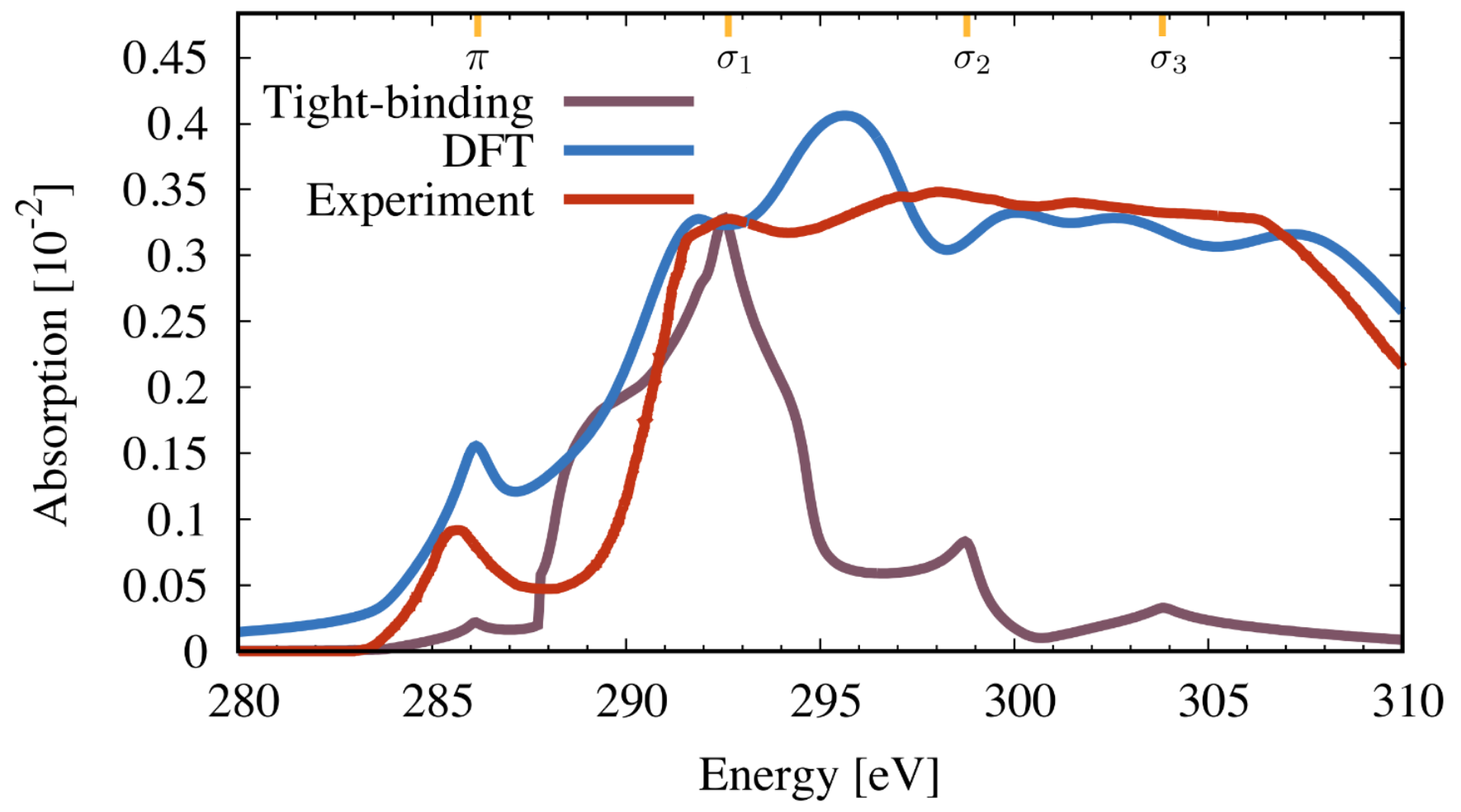}
\end{center}
\caption{The purple curve shows the XANES of graphene with an analytical 5-band tight binding model. In contrast, the blue curve takes into account a first principle calculation. When comparing with the experiment (red) we find a dramatically improved match. All curves are normalized to DFT curve.}
\label{fig:xanes}
\end{figure}

In the following, we calculate the X-ray absorption near edge spectrum for an angle of incidence $\theta=16^\circ$. We start by discussing the XANES taking only an analytical 5 band model into account, cf. Fig. \ref{fig:xanes} purple curve. Here, we see the first K-edge rising at \unit[286]{eV} corresponding to the transition of the 1$s$ electron into the $\pi$ band ($2p_z$ orbital). The following the three peaks originate from transition of the 1$s$ core electron into the sigma bands formed by the $2s$, $2p_x$, and $2p_y$ orbitals. Between the three peaks the absorption dramatically decreases. Comparing with experiments, cf. Fig. \ref{fig:xanes} red curve taken from Ref. \cite{Papagno2009}, we observe a qualitatively different behavior. Moreover, we note a plateau corresponding to an almost continuous absorption. Therefore, the analytical five-band tight-binding model is clearly not sufficient for the X-ray absorption near edge spectrum modeling leading us to use the first principles approach. The \textit{ab initio} band structure calculations enable us to include a large number of electronic bands with exceptional accuracy across the entire Brillouin zone (Fig. \ref{fig:bands}). The blue curve in Fig. \ref{fig:xanes} shows the obtained XANES using the first principles electron bands. We see that in addition to the peaks at \unit[286]{eV} and \unit[292]{eV}, we recover now also the plateau starting at \unit[292]{eV} after a rapid increase of the absorption edge. Furthermore, the decreasing absorption from \unit[307]{eV} is well described. However, the absorption around \unit[296]{eV}  computed in our work exceeds the values observed in the experiment. This shows that the oscillator strength of the transitions into higher sigma bands are overestimated in our calculation. In general, the oscillator strength is slightly overestimated, when observing the $\pi$ peak (\unit[286]{eV}). However, the peak height ratio of the $\pi$ peak and the highest $\sigma$ peak corresponds well to the ratio observed in the experiment. Additionally, so far no microscopic description of the band-dependent dephasing is present. An accurate derivation does not only give valuable insights into the relaxation of core-X-ray excitations in two-dimensional materials but also leads to a proper distribution of the oscillator strength over all bands.

\section{Conclusion}

In summary, we employed the \textit{ab initio} calculations applied to graphene, using a pseudopotential, which contains information about the outer core shells (1$s$) and obtained the graphene band structure and Bloch factors including the 1$s$ core shell from the solution of the Kohn-Sham equation as implemented by \textsc{Quantum ESPRESSO}. The Bloch factors allowed for a calculation of the K-edge dipole elements in dipole approximation as implemented in YAMBO. The first-principle results were used as input for the microscopic Maxwell-X-ray Bloch formalism for X-ray absorption. This leaded to an accurate XANES of graphene also at high energies as compared to a semi-empirical tight-binding approach. 

Our work shows that on one hand the many-band scenario appearing in the microscopic Maxwell-X-ray Bloch formalism can comfortably be evaluated by including the results from first-principle calculations which leads to much better agreement with experiment than an analytical few-band model. On the other hand, first-principle codes, especially such that focus on core-electron spectroscopy as OCEAN \cite{vinson2011bethe,gilmore2015efficient}, can implement the electric field and susceptibility tensor developed in this work to calculate the real absorption as a macroscopic observable instead of the joint density of states or Fermi's golden rule for the transition amplitude.

\begin{acknowledgments}
We thank Dhruv Desai (Caltech) and Yao Luo (Caltech) for fruitful discussions. We acknowledge financial support from the Deutsche Forschungsgemeinschaft (DFG) through KN427 14-1, Theory of core-exciton dynamics and nonlinear X-ray spectroscopy (project number 527838492) (D.C., J.S., A.K.). I.M. and M.B. acknowledge support by the Liquid Sunlight Alliance, which is supported by the U.S. Department of Energy, Office of Science, Office of Basic Energy Sciences, under Award Number DE-SC0021266.
\end{acknowledgments}

\newpage
\appendix

\section{Green's dyade from the wave equation}

We start from the wave equation 
\begin{align}
    \nabla^2\mathbf{E}(\mathbf{r},t)-\frac{\epsilon}{c^2}\frac{\partial^2}{\partial t^2}\mathbf{E}(\mathbf{r},t)&=\mu_0\frac{\partial^2}{\partial t^2}\mathbf{P}(\mathbf{r},t) \nonumber \\
    &-\frac{1}{\epsilon\epsilon_0}\nabla(\nabla\cdot\mathbf{P}(\mathbf{r},t)) \;,
\end{align}
which can be derived from the Maxwell equations. Since we want to include non-local phenomena and a non-perpendicular excitation geometry the second term on the right hand side does not vanish. The wave equation can be solved with a Green's function approach, which reads in Fourier space
\begin{align}
    \mathbf{E}_{\mathbf{Q}}(Q_z,\omega)=G_{\mathbf{Q}}(Q_z,\omega)\boldsymbol{\Pi}_{\mathbf{Q}}(Q_z,\omega)
\end{align}
with the inhomogeneity $\boldsymbol{\Pi}_{\mathbf{Q}}(Q_z,\omega)$ corresponding to the right hand side of the wave equation. By Fourier transformation of the wave equation in space and time allows for an identification of the Green's function. A subsequent inverse Fourier transformation with respect to $Q_z$ yields the solved electric field
\begin{align}
    \mathbf{E}_{\mathbf{Q}}(z,\omega)=\int dz'~G_{\mathbf{Q}}(z-z',\omega)\boldsymbol{\Pi}_{\mathbf{Q}}(z',\omega) \label{eq:Efield-integral}
\end{align}
with $G_{\mathbf{Q}}(z-z',\omega)=i\exp(-i\kappa |z-z'|)/2\kappa$ and $\kappa^2=\epsilon\omega^2/c^2-|\mathbf{Q}|^2$. Assuming that the macroscopic polarization $\mathbf{P}_{\mathbf{Q}}(z,\omega)$ (defining $\boldsymbol{\Pi}_{\mathbf{Q}}(Q_z,\omega)$) is located in-plane of the sample, i.e. $\mathbf{P}_{\mathbf{Q}}(z,\omega)=P^{2D}_{\mathbf{Q}}(\omega)\delta(z-z')$, the integral in Eq. \eqref{eq:Efield-integral} can be solved. Writing the solution in matrix form $\mathbf{E}_{\mathbf{Q}}(z,\omega)=[G_{\mathbf{Q}}](z,\omega)\mathbf{P}_{\mathbf{Q}}^{2D}(\omega)$ we can define the Green's dyade from the main text as
\begin{widetext}
\begin{align}
    [G_{\mathbf{Q}}](z,\omega)&=\frac{e^{-i\kappa |z-z_0|}}{2i}\begin{pmatrix}
\frac{\mu_0\omega^2}{\kappa}-\frac{Q_x^2}{\epsilon_0\epsilon\kappa} & -\frac{Q_xQ_y}{\epsilon_0\epsilon\kappa} & \frac{Q_x}{\epsilon_0\epsilon}\text{sgn}(z-z_0) \\
-\frac{Q_yQ_x}{\epsilon_0\epsilon\kappa} & \frac{\mu_0\omega^2}{\kappa}-\frac{Q_y^2}{\epsilon_0\epsilon\kappa} & \frac{Q_y}{\epsilon_0\epsilon}\text{sgn}(z-z_0) \\
\frac{Q_x}{\epsilon_0\epsilon}\text{sgn}(z-z_0) & \frac{Q_y}{\epsilon_0\epsilon}\text{sgn}(z-z_0) & \frac{\mu_0\omega^2}{\kappa}-\frac{\kappa}{\epsilon_0\epsilon} \;.
\end{pmatrix}
\end{align}
\end{widetext}

\section{Bl\"ochl formalism}
 The Bl\"ochl formalism \cite{blochl1994projector} is based on the linear transformation $\mathcal{T}$ between the true wave function $\psi_{n,\mathbf{k}}$ and an auxiliary wave function $\tilde{\psi}_{n,\mathbf{k}}$: $\psi_{n,\mathbf{k}}=\mathcal{T}\tilde{\psi}_{n,\mathbf{k}}$. The space is divided into atom-centered augmentation spheres and bonding region outside the sphere. The wave function matches at the boundary of the spheres. Since beyond the critical core radius the wave function is already smooth, the transformation attacks only inside the core sphere. Here, the true wave function is expanded in partial waves $\phi^a_i$, which them self are expressed by smooth auxiliary functions $\tilde{\phi}^a_i$ with atom index $a$ and compound index $i$ for the atomic site and angular quantum numbers. The all-electron Kohn-Sham wave functions are described by
\begin{align}
\psi_{n,\mathbf{k}}(\mathbf{r})=\tilde{\psi}_{n,\mathbf{k}}(\mathbf{r})+\sum_{a,i}\left(\phi^a_i(\mathbf{r})-\tilde{\phi}^a_i(\mathbf{r})\right)\langle\tilde{p}^a_i|\tilde{\psi}_{n,\mathbf{k}}\rangle \;. \label{eq:Blochltrafo}
\end{align}
With the linear transformation defined in Eq. \eqref{eq:Blochltrafo}, the smooth auxiliary wave function $\tilde{\psi}_{n,\mathbf{k}}$ are obtained by solving the transformed Kohn-Sham equation and expressed in a plane wave basis set. In contrast, $\phi^a_i$, $\tilde{\phi}_i^a$, and $\tilde{p}_i^a$ are atom-centered localized functions. $\phi^a_i(\mathbf{r})$ are all-electron partial wave solutions to the radial scalar relativistic non spin-polarized Schr\"odinger equation and multiplied by the spherical harmonics. A pseudopotential procedure yields $\tilde{\phi}_i^a$ and $\langle\tilde{p}_i|\tilde{\phi}_j\rangle=\delta_{i,j}$ with projector functions $\tilde{p}^a_i$ probing the character of the wave function. The set of functions are system independent. Therefore, they are pre-calculated in an atomic environment, tabulated, and kept frozen in the remaining calculation.


\end{document}